# The impact of climate change on astronomical observations

*Climate change is affecting and will increasingly affect astronomical observations, particularly in terms of dome seeing, surface layer turbulence, atmospheric water vapour content and the wind-driven halo effect in exoplanet direct imaging.*

**Faustine Cantalloube[1], Julien Milli[2,3], Christoph Böhm[4], Susanne Crewell[4], Julio Navarrete[3], Kira Rehfeld[5], Marc Sarazin[6] and Anna Sommani[5]**

[1]Max Planck Institute for Astronomy, Königstuhl 17, 69117 Heidelberg, Germany
[2]Université Grenoble Alpes, CNRS, IPAG, 38000 Grenoble, France
[3]European Southern Observatory (ESO), Alonso de Córdova 3107, Vitacura, Casilla 19001, Santiago, Chile
[4]Institute for Geophysics and Meteorology, University of Cologne, Cologne, Germany
[5]Institute of Environmental Physics and Heidelberg Center for the Environment, Heidelberg University, INF229, 69120 Heidelberg, Germany
[6]European Southern Observatory, Karl-Schwarzschild-Str. 2, D-85748 Garching, Germany

Corresponding author email: cantalloube@mpia.de

Astronomers are entering an era in which they change the way they will work, with the arrival of 30–40-m class ground-based telescopes and large international observational projects, sparking new ways of communicating and collaborating. These scientific challenges come together with societal ones, such as the role astronomers play in communicating and undertaking actions to significantly reduce the environmental footprint of astronomical research. More generally, it is urgent that astronomers, through their unique perspective on the universe, communicate and act about climate change consequences at any level. In this context, we have investigated the role some key weather parameters play in the quality of astronomical observations, and analysed their long-term trends (longer than 30 years) in order to grasp the impact of climate change on future observations. In the following, we give four examples of how climate change already affects or could potentially affect the operations of an astronomical observatory. This preliminary study is conducted with data from the Very Large Telescope (VLT), operated by the European Southern Observatory (ESO), located at Cerro Paranal in the Atacama desert, Chile, which is one of the driest places on Earth. For the analyses presented below, we used the various sensors installed at the Paranal Observatory but also, to show a longer time span (from 1980 to present), we used the fifth generation European Centre Medium-Range Weather Forecasts (ECMWF) atmospheric reanalysis of the global climate, ERA5 [1], with a spatial resolution of 31km, which we interpolated at the Paranal observatory location. To investigate longer timescale evolution (from 1900 to 2010), at a cost of a coarser spatial resolution (130km) that averages the actual orography and may blend the ocean/continent interfaces, we also used in some cases the ERA20C reanalysis data [2]. In addition, we also explored climate projections in this region, using the Coupled Model Intercomparison Project Phase 6 (CMIP6) multi-model ensemble [3], under the worst-case climate change Shared Socio-Economic Pathways (SSP5-8.5) scenario. Further investigation is needed to better understand the underlying mechanisms of change, as well as to assess the severity of the impact.

The first example is a consequence of the increase in surface temperature locally at the Paranal observatory, which affects any type of observation sensitive to the atmospheric seeing that degrades the spatial resolution of the telescopes. During the day, the temperature inside the dome enclosure of the telescopes is cooled down to correspond to the outside temperature during the dome opening (at sunset). The thermal system used for the four unit telescopes (UTs) is an active control system aimed at minimizing the difference between the telescope temperature and the ambient temperature. At the end of each night, a prediction of the temperature for the following sunset is made from the ECMWF and is set as the target temperature (set point) for the thermal system during the day. Currently, this system cannot reach the target temperature if it is above $16^{o}$C. However, with the increase in temperature at the location of the Paranal observatory by $1.5^{o}$C on average within the last 40 years (Fig.1a), consistent with the anthropogenic global warming [4], there are more and more occurrences of target temperatures exceeding the current cooling system limitations of $16^{o}$C (Fig.1b,c). In such a case, the outside temperature is warmer than the inner dome temperature during opening. The temperature difference between the telescope, particularly the primary mirror, and the ambient air enhances internal turbulence inside the dome: it is called dome seeing [5]. The dome seeing degrades the image quality by causing blurring. In addition, we show the CMIP6 projection model of the surface temperature at the location of the Paranal observatory, using the SSP5-8.5 scenario (Fig. 1d), pointing towards an average increase of another $4^{o}$C at the end of the century. This transition should be taken into consideration for the on-going construction of the Extremely Large Telescope (ELT) at the Cerro Armazones situated 20km eastward from the Cerro Paranal, and for the design of its instruments that are currently under development.

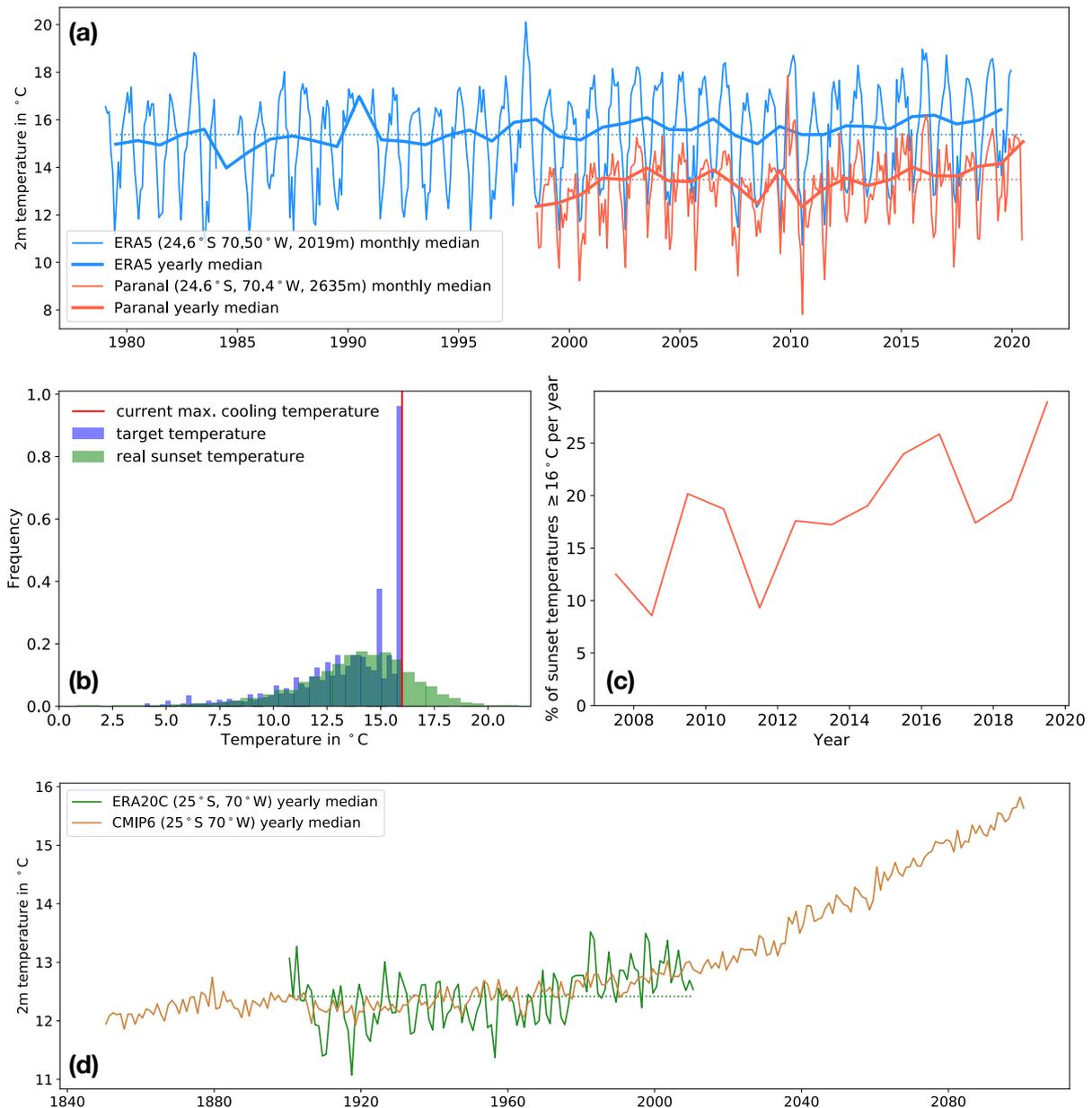

**Figure 1 | Temperature in the region around Paranal Observatory.** (a) Monthly averaged daily mean temperature over the Paranal observatory as a function of time, retrieved from the ERA5 reanalysis data (blue) and as measured at the Paranal observatory (red), with the corresponding yearly average (thick lines) and median (dashed lines). (b) Occurrence of the real (green) and target (blue) temperature (limited to 16$^o$C, red solid line) of the UTs dome cooling system, from 2006 to 2020. (c) Frequency of the sunset temperature measured at Paranal to be above the 16$^o$C limit of the current cooling system, as a function of time. (d) Yearly median near surface air temperature as a function of time, from the ERA20C reanalysis data (green) with its global median (green dotted line), and from the CMIP6 climate projection using the SSP5-8.5 scenario (Beijing Climate Centre, BCC-CSM2-MR model ensemble), adjusted to the ERA20C mean (orange).

The second example is the increase in the surface layer turbulence, leading to stronger seeing measured in recent years (Fig. 2). The surface layer is a thin, time-variable layer located in the first tens of meters above ground, which contributes to a large fraction of the optical turbulence due to the inefficient heat exchange between the ground and the airflow [6]. However, the UTs are sensitive to the turbulence from 10 to 30m on (depending on the wind conditions and pointing direction). And indeed, the image quality, as measured directly at the UTs, is not affected by the increase in surface layer so far and is rather constant [7]. At the Cerro Paranal, the seeing is measured by a Differential Image Motion Monitor (DIMM, [8]). In Fig.2, we show a time series of seeing measurements from 1986 to 2020. The DIMM height above ground has changed through time: in 1986 it was installed on the mountaintop, 28m higher that the current platform; in 1992 the mountaintop was levelled and it was installed on a 5-m tower; from 1994 to 2000 the four UTs were built; and in 2003 the VLT survey telescope (VST) was built. In 2016, a second DIMM was mounted on a 7-m tower, at a location where it is less impacted by disturbances due to the UTs or the VST. To match the two DIMM measurements, the measurements of the older DIMM were adjusted to the mean value of the new one. The measured increase in seeing (Fig. 2) is due to increased surface layer turbulence [7]. The origin of this increase can be mainly explained by two hypotheses: either (i) the levelling of the mountain and the numerous changes of configuration of the DIMM altitude and the remodelling of the air fluxes due to the erection of various buildings, or (ii) the local changes due to global atmospheric circulation transition under on-going climate change [9]. The first hypothesis is supported by the fact that the new DIMM finds similar seeing values as the first DIMM before the levelling of the mountain. The second hypothesis is supported by the clear increase in the surface temperature (Fig. 1a), potentially leading to higher temperature gradients and therefore higher radiative cooling or convection, which induces stronger turbulence close to the ground. Computational fluid dynamics simulations of the turbulence at the Paranal site or extensive understanding of the global to local scale mechanisms would allow us to address this question.

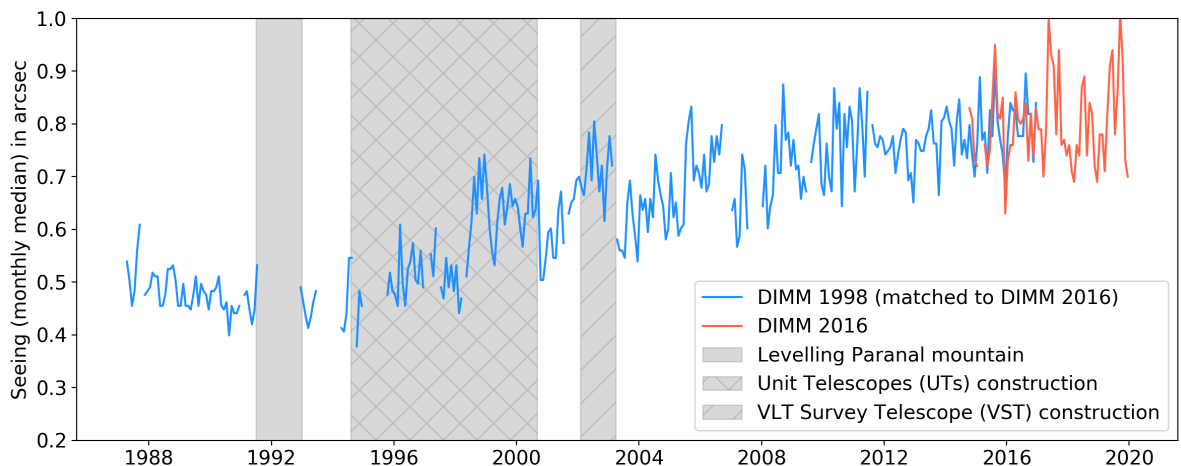

**Figure 2 | Surface layer seeing.** Evolution of the seeing measured by the two DIMM instruments (blue and red lines) installed at Cerro Paranal from 1986 (before flattening the platform in 1991) to today. The blue line is normalized to match the red line.

The third example is in the context of the demanding technique of exoplanet imaging, requiring both a very high angular resolution and a high contrast (about $10^{-6}$ at 500 milliarcseconds in the near-infrared). Adaptive optics correct for the atmospheric turbulence in near real time and provide an angular resolution close to the theoretical diffraction of the 8-m telescope. This enables the use of a coronagraph to reach a high-contrast regime. However, the time lag between the analysis of the atmospheric turbulence and its correction by a deformable mirror creates a wind-driven halo (Fig. 3a). The southern subtropical jet stream (located at about 12km altitude, with wind speeds up to 60m s$^{-1}$) is primarily responsible for the wind-driven halo [10,11]. This structure arises 30 to 40% of the time and leads to a tenfold reduction in contrast [10]. We therefore looked at the evolution of the horizontal wind speed at the jet stream layer (200mbar) over time from the ERA5 reanalysis data (Fig. 3b). On the monthly average, we can directly see the seasonal changes (being more prominent in winter) and on the yearly average we see longer trends linked with the El Niño Southern Oscillation (ENSO, 2 to 7 years variation) and the Pacific Decadal Oscillation (PDO, on longer time scales), both linked to the sea surface temperature (SST) anomalies. Over the 40 years of ERA5 data, we observe a slight increase in the average wind speed by about 3–4 m s$^{-1}$. When relating to the Niño 3.4 index [12] (the best proxy for the strength of the ENSO), we observe a large fraction of wind-driven halo during strong El Niño events (the warm phase of the ENSO), such as the one in 2015. Recent modelling studies suggest that, due to greenhouse-effect-related warming, the ENSO variability will increase and El Niño will intensify [13,14]. In a further study, we will gather astronomical data sensitive to the wind-driven halo [10] and will correlate it to the ERA5 data and meso-scale atmospheric models to further investigate the evolution of the jet stream over the Paranal observatory.

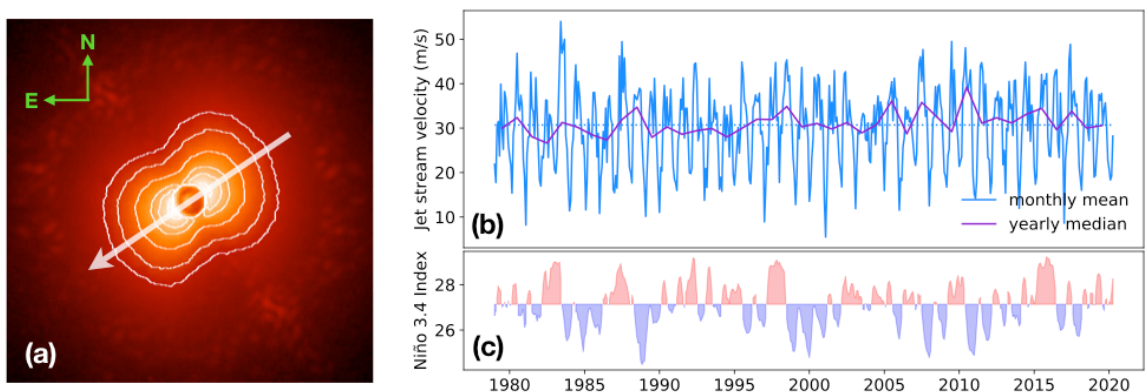

**Figure 3 | Horizontal speed of the jet stream.** (a) Wind-driven halo observed in a VLT/SPHERE coronagraphic image. The wind direction is indicated with the white arrow. (b) Monthly averaged horizontal wind speed at the jet stream layer (200mbar) as a function of time from the ERA5 reanalysis data. (c) Niño 3.4 index showing El Niño (red) and La Niña (purple) events as a function of time. El Niño or La Niña events are defined when the Niño 3.4 sea surface temperatures anomaly, filtered with a five-month running mean, exceed ±0.4 °C for a period of six months or more.

Finally, three critical parameters affecting the scheduling and availability of astronomical instruments at large observatories are the integrated water vapour (IWV), the relative humidity, and the cloud coverage. Atacama is the driest place on Earth after Antarctica but still experiences high humidity events such as yearly altiplanic winters (also called South American summer monsoon, [15] occurring in January/February and visible in Fig.4a) and episodic atmospheric rivers (two to three days of high humidity [16]). Strong IWV events are primarily related to high central equatorial sea surface temperature during El Niño events. Recent climate studies give a hint that an increased $CO_2$ concentration in the atmosphere will give rise to a global increase in humidity [17] and to more violent El Niño events [13], which means more frequent severe flooding in the Atacama region, as witnessed in March 2015 [18]. Here, we extracted the IWV at the Paranal observatory location from the ERA5 reanalysis, complemented by the ERA20C reanalysis that covers a full century (to probe the 4-year ENSO cycle) [19], which we compared to on-site measurements using the Low Humidity and Atmospheric Temperature PROfiler (LHATPRO), a radiometer installed at the Paranal observatory [20] (Fig. 4a,b). We do not observe an obvious trend in the mean, extremes of IWV or in the frequency or duration of high humidity events, indicating a long-term effect due to climate change. The notable trend of increasing low IWV events revealed by the radiometer over the last 5 years (Fig.4b) is not over a long enough timescale to be able to draw any statistical conclusions. By exploring different projection models, there are hints that the area will become drier, but it needs to be more thoroughly checked due to the coarse resolution of the current models, added to the very specific location of the Paranal observatory, at the interface between the Andes and the Chilean coastal range. In the future, we will refine this analysis to better apprehend the potential variation of IWV in the context of large near-infrared, submillimetre and radio wavelengths observatories.

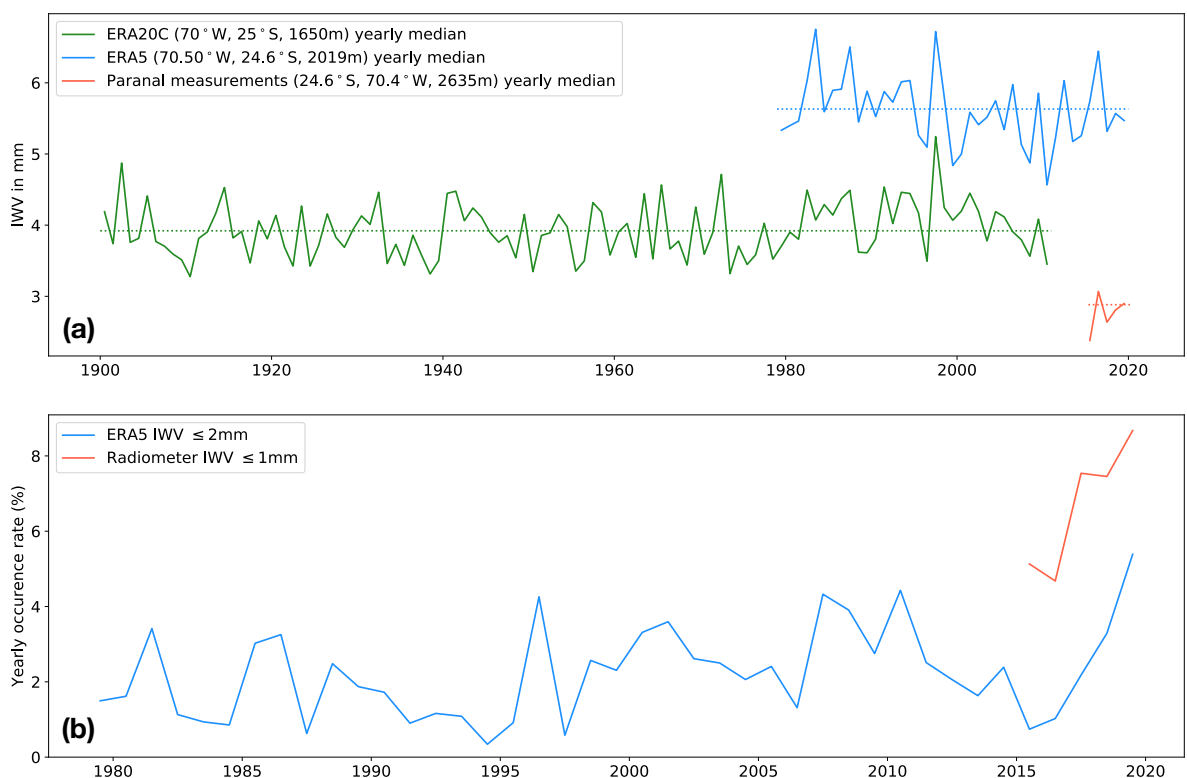

**Figure 4 | Integrated water vapour.** (a) Yearly averaged integrated water vapour (IWV) as a function of time from the ERA20C reanalysis data (green), ERA5 reanalysis data (blue) and the radiometer

measurement at Paranal observatory (red). The offset between the three curves is due to the poor spatial resolution of the models used for the reanalysis (130km for ERA20C and 31km for ERA5) and the difference in altitude (1650m, 2019m and 2635m, respectively). (b) Occurrence of the IWV to be below a threshold of 2mm for the ERA5 reanalysis data (blue line) and below 1mm for the LHATPRO data (red line). Because the IWV estimated from the ERA5 reanalysis data is twice higher than the Paranal real measurements, we used a threshold twice higher for ERA5 to get comparable occurrence rates of low IWV events. While the absolute values can only be trusted from the measurements, the trend can be compared to reanalysis.

The purpose of this work is to raise awareness among the astronomy community about the effect and immediate consequences of climate change on astronomical data. Each telescope site is likely to have its own microclimate that calls for individual study, but we have highlighted three different areas — dome seeing, surface layer turbulence and the wind-driven halo effect – that will have an increasingly detrimental impact on astronomical observations at Paranal as climate change worsens. Increasing atmospheric water content in some areas may also degrade observations, but for Paranal at least, there are indications the region might get drier. Through this work we have intensified collaborations between astronomers, climatologists, atmospheric scientists and desert area specialists, and we encourage others to do the same. The measurements of weather quantities in the southern hemisphere are scarce, not only due to the smaller amount of landmass, but also due to the economic status of the Southern countries (for example, fewer airports means less data are collected) and satellite data are too recent for long-trend analysis. For more than 30 years, astronomical observatories have been collecting daily weather data that can be jointly used to study the effect of climate change and refine the link between synoptic and optical turbulence scales, in particular in Chile where interactions between the ocean, the coastal area and the Andes involve complex mechanisms.

As astronomers, we are privileged to work in such a fascinating field, studying objects beyond Earth for the sole purpose of increasing humanity's knowledge. Our work consists of thinking outside of the box, solving problems and analysing data with critical thinking. On top of this, our subject of study goes beyond any political or financial stakes. As a consequence, we have the opportunity, the tools and the prospect to build and follow concrete and sustainable actions against climate crises such as (i) communicating inside and outside our community about the impact of climate change on our planet and our society, (ii) optimising the energy resources expended for our professional activity, and (iii) revisiting and reshaping the whole scheme of research in astronomy to decrease our global footprint. To do so, a massive cultural shift is needed, and it is of prime importance that astronomy uses its unique perspective to claim this simple fact: there is no planet B.

**Acknowledgements**

We would like to thank our collaborators who assisted us in collecting the data and sharing their thoughts about the present study: Maxime Boccas (ESO), Eloy Fuenteseca (ESO), Iván Muñoz (ESO), Eduardo Peña (ESO), Elena Masciadri (INAF), Florian Kerber (ESO), Saavidra Perera (MPIA), Leonardo Burtscher (NOVA) and the *Astronomers For Planet Earth* collective. Christoph Böhm was supported by the Deutsche Forschungsgemeinschaft (DFG, German Research Foundation) – project number 268236062 – SFB 1211. We acknowledge the World Climate Research Programme, which, through its Working Group on Coupled Modelling, coordinated and promoted CMIP6. We thank the climate modelling groups for producing and making available their model output, the Earth System Grid Federation (ESGF) for archiving the data and providing access, and the multiple funding agencies who support CMIP6 and ESGF.

**Competing Interests**
The authors declare no competing interests.